\documentclass[conference]{IEEEtran}
\usepackage{epsfig,cite,bm}
\def\oc{\mathop{\mbox{\scriptsize (OC)}}}
\def\toc{\mathop{\mbox{\tiny (OC)}}}
\def\mrc{\mathop{\mbox{\scriptsize (MRC)}}}
\def\tmrc{\mathop{\mbox{\tiny (MRC)}}}
\def\sc{\mathop{\mbox{\scriptsize (SC)}}}
\def\tsc{\mathop{\mbox{\tiny (SC)}}}

\IEEEoverridecommandlockouts
\begin{document}
\title{SINR Analysis of Opportunistic MIMO-SDMA Downlink Systems with Linear Combining}
\author{Man-On Pun, Visa Koivunen and H. Vincent Poor
\thanks{Man-On Pun and H. Vincent Poor are with the Department of Electrical Engineering, Princeton University, Princeton, NJ 08544.}
\thanks{Visa Koivunen is with the Signal Processing Laboratory, Helsinki University of Technology (HUT), Finland.}
\thanks{This research was supported in part by the Croucher Foundation under a post-doctoral fellowship, and in part by the U. S. National Science Foundation under Grants ANI-03-38807 and CNS-06-25637.}}
\maketitle
\thispagestyle{empty}

\begin{abstract}
Opportunistic scheduling (OS) schemes have been proposed previously by the authors for multiuser MIMO-SDMA downlink systems with linear combining. In particular, it has been demonstrated that significant performance improvement can be achieved by incorporating low-complexity linear combining techniques into the design of OS schemes for MIMO-SDMA. However, this previous analysis was performed based on the effective signal-to-interference ratio (SIR), assuming an interference-limited scenario, which is typically a valid assumption in SDMA-based systems. It was shown that the limiting distribution of the effective SIR is of the Frechet type. Surprisingly, the corresponding scaling laws were found to follow $\epsilon\log K$ with $0<\epsilon<1$, rather than the conventional $\log\log K$ form.

Inspired by this difference between the scaling law forms, in this paper a systematic approach is developed to derive asymptotic throughput and scaling laws based on signal-to-interference-noise ratio (SINR) by utilizing extreme value theory. The convergence of the limiting distribution of the effective SINR to the Gumbel type is established. The resulting scaling law is found to be governed by the conventional $\log\log K$ form. These novel results are validated by simulation results. The comparison of SIR and SINR-based analysis suggests that the SIR-based analysis is more computationally efficient for SDMA-based systems and it captures the asymptotic system performance with higher fidelity.
\end{abstract}

\section{Introduction}\label{sec:intro}
Opportunistic scheduling (OS) has recently attracted considerable research interest as a promising technique to improve system throughput by exploiting multi-user diversity with limited channel feedback \cite{Tse02}. Generally speaking, existing OS schemes can be classified into two categories, namely the time-sharing (TS) \cite{Tse02} and space-division multiple access-based (SDMA)-based \cite{Sharif05} OS schemes. In TS-OS, only the mobile terminal (MT) with the best instantaneous channel conditions is scheduled in one slot, regardless of the number of beams employed by the base station (BS). In contrast, SDMA-based OS serves multiple MTs {\em simultaneously} with multiple orthonormal beams in each slot. Denote by $M$ and $N$ the number of transmit and receive antennas, respectively. It has been shown recently  that the sum-rate of SDMA-based OS grows linearly with $M$ whereas that of TS-OS increases only linearly with $\min(M,N)$ \cite{Sharif07}. In addition to the more rapidly growing scaling law, SDMA-based OS is particularly attractive for practical systems with stringent latency requirements.

The SDMA-based OS in \cite{Sharif05} was originally developed for systems with single-antenna MTs. For MTs with multiple receive antennas, \cite{Sharif05} proposes to let each antenna compete for its desired beam as if it were an individual MT. As a result, each beam is assigned to a specific receive antenna of a chosen MT. Since signals received from the undesignated antennas of a chosen MT are discarded, this leads to inefficient utilization of multiple receive antennas. In \cite{Pun07}, various linear combining techniques exploiting signals received by all receive antennas were proposed. The enhanced effective SINR is employed as a scheduling metric. Both analytical and simulation results in \cite{Pun07} have demonstrated that the system sum-rate performance can be significantly improved by using such combining techniques. For instance, the optimal combining technique can provide over $40\%$ sum-rate improvement compared to the selection combining technique for $M=4$ and $N=2$ \cite{Pun07}.

The theoretical analysis in \cite{Pun07} has been conducted based on SIR, assuming an interference-limited environment. The resulting scaling laws have a distinctive form, i.e. $\epsilon\log K$ with $0<\epsilon<1$, which is very different from the conventional form  $\log\log K$ derived based on signal-to-noise-ratio (SNR) \cite{Tse02} or SINR \cite{Sharif05,Sharif07} in the literature. Similar results have been independently developed for multicell systems in \cite{Gesbert07}. In this work, we introduce a systematic approach for deriving asymptotic throughput and scaling laws using SINR. The proposed approach stems from extreme value theory\cite{Gumbel58}. We prove that the cumulative distribution functions (CDFs) of the effective SINR obtained with linear combining converge to the Gumbel-type limiting distribution. Furthermore, we show that the SINR-based scaling laws for the proposed opportunistic beamforming and scheduling schemes follow the conventional $\log\log K$ form. Through comparison between the SIR and SINR-based analysis, it is argued that the SIR-based analysis is more computationally efficient for SDMA-based systems, and subsequently more effective in capturing the high-order behavior of the asymptotic system performance. To make comparison with our previous SIR-based analysis reported in \cite{Pun07}, we concentrate on a practical system with $M=4$ and $N=2$ in this work. However, it should be emphasized that the analysis can be easily generalized for systems with arbitrary $M$ and $N$.

\underline{Notation}: Vectors and matrices are denoted by boldface letters.  $\left\|\cdot\right\|$ represents the Euclidean norm of the enclosed vector  and $\left|\cdot\right|$ denotes the amplitude of the enclosed complex-valued quantity. ${\bm I}_N$ is the $N\times N$ identity matrix. We use $E\left\{\cdot\right\}$ for expectation.  Finally,  $\log$ and $\ln$ are the logarithms to the base $2$ and $e$, respectively.

\section{Signal Model}\label{sec:smodel}

\begin{figure}[htp]
\begin{center}
\includegraphics[scale=0.55]{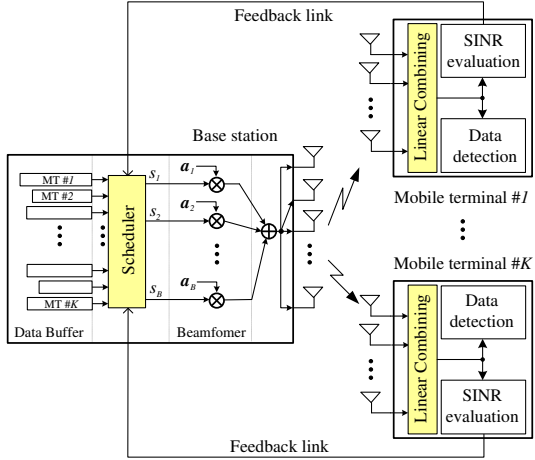}
\caption{A block diagram of the opportunistic MIMO SDMA downlink system under consideration.}\label{fig:system}
\end{center}
\end{figure}

We consider the opportunistic MIMO-SDMA downlink system depicted in Fig.~\ref{fig:system} where the BS is equipped with $M$ transmit antennas and each of the $K$ MTs has $N$ receive antennas with $N\leq M$. Let $\left\{{\bm a}_m;m=1,2,\cdots,M\right\}$ be a vector set containing $M$ orthornormal beamforming vectors of length $M$. We focus on a particular time slot during which a beamforming vector set $\left\{\bm a_m\right\}$ has been chosen from a common codebook shared by the BS and MTs. During the $p$-th slot, the transmitted signal can be expressed as
\begin{equation}
{\bm x}(p)=\sum_{m=1}^M{\bm a}_ms_m(p)={\bm A}{\bm s}(p),
\end{equation}
where ${\bm A}=\left[{\bm a}_1,{\bm a}_2,\cdots,{\bm a}_M\right]$ is the unitary beamforming matrix with ${\bm A}^H{\bm A}={\bm I}_M$ and ${\bm s}(p)=\left[s_1(p),s_2(p),\cdots,s_M(p)\right]^T$ with $E\left\{\left|s_m(p)\right|^2\right\}=1$ is the data vector transmitted in the $p$-th slot. The corresponding received signal by the $k$-th MT can be written as
\begin{equation}
{\bm y}_k(p)=\sqrt{\rho_k}{\bm H}_k(p){\bm x}(p)+{\bm n}_k(p),\label{eq:vyk}
\end{equation}
where ${\bm H}_k$ is the channel gain matrix between the BS and the $k$-th MT with independent and identically-distributed (i.i.d.) Rayleigh-distributed complex  entries. Furthermore, the noise term ${\bm n}_k(p)$ is modeled as ${\cal CN}\left({\bm 0},{\bm I}_N\right)$ and $\rho_k$ is a constant related to the average received SNR given by $E\left\{\rho_k\left\|{\bm H}_k(p){\bm x}(p)\right\|^2\right\}=\rho_kM$.

To keep our following analysis tractable, we concentrate on a homogenous system with $\rho_k=\rho$ in this work. For notational simplicity, we drop the temporal index $p$ in the sequel. Furthermore, we refer to the SINR obtained by linearly combining signals from all receive antennas as the {\em effective} SINR in order to distinguish it from the {\em observed} SINR without combining.

\section{OS with Linear Combining}
In this section, we briefly review the beamforming and scheduling schemes for MIMO-SDMA systems with linear combining techniques proposed in \cite{Pun07}. As shown in Fig. \ref{fig:system}, in the beginning of a time slot, each MT evaluates the effective SINR for each beam by linearly combining the received signals with one of the following three combining techniques, namely selection combining (SC), maximum ratio combining (MRC) and optimum combining (OC) before returning the information about $M$ effective SINRs to the BS. Note that OC performs active interference suppression by exploiting the interference structure, whereas MRC and SC simply intend to amplify the desired signal. It will be shown later that this characteristic interference-suppression feature of OC enables the scheduling scheme with OC to considerably outperform those with SC and MRC.

Upon receiving the effective SINR information from all MTs, the BS schedules and starts data transmission to multiple MTs with the largest effective SINRs on different beams until the end of the current time slot. At each chosen MT, received signals from all antennas are linearly combined using one of the above linear combining techniques, followed by data detection. It is worth noting that the probability of awarding multiple beams to the same MT is rather small, as the number of MTs is large. Furthermore, recall that the minimum mean squared error (MMSE) and zero-forcing (ZF) receiver structures for MIMO receivers amount to combiners using OC and MRC for each beam, respectively. As a result, for an MT assigned with multiple beams, it can focus on one assigned beam at a time using the chosen combining technique while regarding all other beams as interfering sources.

\section{SINR Analysis}\label{sec:analysis}
Define $\gamma^*_m=\max\left(\gamma_{1,m},\gamma_{2,m},\cdots,\gamma_{K,m}\right)$, for $m=1,2,\cdots,M$. Assuming $\gamma_{k,m}$ for $k=1,2,\cdots,K$, are i.i.d. with CDF $F_X(x)$, the resulting average system throughput can be computed as \cite{Sharif05}:
\begin{eqnarray}
C&=&E\left\{\sum_{m=1}^M\log\left(1+\gamma^*_m\right)\right\}\nonumber\\
&=&M\int_0^\infty\log\left(1+x\right)\,d\left[F_{X}(x)\right]^K.\label{eq:sumrate}
\end{eqnarray}

In the following, we first derive $F_{X}(x)$ based on different linear combining techniques before establishing their corresponding limiting distributions, i.e. $\displaystyle\lim_{K\rightarrow\infty}\left[F_{X}(x)\right]^K$. By exploiting the limiting distributions, we derive the asymptotic throughput and the corresponding scaling laws. In the sequel, we focus on a practical system with $M=4$ and $N=2$. However, it has been shown in \cite{PunJSAC07} that the analysis can be easily generalized for systems with arbitrary $M$ and $N$.

\subsection{Selection Combining (SC)}
We begin with the selection combining. Denote by $x$ the maximum of the two SINR values of the $i$-th beam perceived by the two antennas at the $k$-th MT. The CDF of $x$ can be derived based on the results in \cite{Sharif05} and reads
\begin{equation}\label{eq:sccdf}
F^{\sc}_X(x)=\left[1-\frac{e^{-x/\rho}}{\left(1+x\right)^3}\right]^2.
\end{equation}

Differentiating $F^{\sc}_X(x)$ with respect to $x$, we can obtain the corresponding probability density function (PDF).
\begin{equation}\label{eq:scpdf}
f^{\sc}_X(x)=2\left[1-\frac{e^{-x/\rho}}{\left(1+x\right)^3}\right]\frac{(1+x)\frac{1}{\rho}e^{-x/\rho}+3e^{-x/\rho}}{(1+x)^4}.
\end{equation}

It is straightforward to show that $F^{\sc}_X(x)$ and $f^{\sc}_X(x)$ satisfy the following equation
\begin{equation}
\lim_{x\rightarrow\infty}\frac{1-F^{\sc}_X(x)}{f^{\sc}_X(x)}=\rho>0,
\end{equation}
which is the necessary and sufficient condition for the limiting distribution of $\left[F^{\sc}_X(x)\right]^K$
being of the Gumbel type \cite{Gumbel58}. Consequently, $F_{X_{(K)}}(x)=\left[F_X(x)\right]^K$ converges to the following Gumbel-type distribution \cite{Gumbel58}.
\begin{equation}\label{eq:gumbel}
F^{\sc}_{X_{(K)}}\left(a^{\tsc}_Kx+b^{\tsc}_K\right)=e^{-e^{-x}},\quad x\geq 0
\end{equation}
or equivalently,
\begin{equation}
F^{\sc}_{X_{(K)}}\left(x\right)= e^{-e^{-\frac{x}{a^{\tsc}_K}+\frac{b^{\tsc}_K}{a^{\tsc}_K}}}, \quad x\geq 0,\label{eq:eqPrM4N2sc}
\end{equation}
where $a^{\tsc}_K$ and $b^{\tsc}_K$ are normalizing factors affecting the shape and location of the limiting distribution, respectively. From extreme value theory, $b^{\sc}_K$ can be computed from the characteristic extreme of (\ref{eq:sccdf}) as \cite{Gumbel58}
\begin{equation}
1-F_X^{\sc}\left(b^{\tsc}_K\right)=\frac{1}{K}.\label{eq:scbk}
\end{equation}
Since (\ref{eq:scbk}) is an exponential-linear equation of $b^{\tsc}_K$, it is non-trivial to obtain the exact solution of $b^{\tsc}_K$ in closed form. Fortunately, since $1-F^{\sc}_{X_{(K)}}$ monotonically decreases from $1$ to $0$ whereas $1/K\in\left[1,0\right)$ for $K=1,2,\cdots,\infty$, there always exists a unique solution of (\ref{eq:scbk}). Thus, we can resort to numerical methods to compute the numerical solution of $b^{\tsc}_K$. It should be emphasized that $1-F_X^{\sc}\left(b^{\tsc}_K\right)$ tends to $1$ as $K$ approaches infinity, which implies that $b^{\sc}_K$ increases with $K$.

Furthermore, $a^{\tsc}_K$ can be obtained from solving the following equation.
\begin{eqnarray}\label{eq:scaK}
a^{\tsc}_K={F_X^{\sc}}^{-1}\left(1-\frac{1}{Ke}\right)-b^{\tsc}_K.
\end{eqnarray}
Similar to $b^{\tsc}_K$, we can show that there always exists a unique solution of $a^{\tsc}_K$. Therefore, the numerical solution of $a^{\tsc}_K$ can be found by resorting to numerical methods.

Finally, the throughput obtained with SC can be computed by substituting (\ref{eq:eqPrM4N2sc}) into (\ref{eq:sumrate}) and reads
\begin{equation}
C^{\sc}=\frac{4}{\ln2}\int_{0}^{\infty}\frac{1-e^{-e^{-\frac{x}{a^{\tsc}_K}+\frac{b^{\tsc}_K}{a^{\tsc}_K}}}}{1+x}\,dx.\label{eq:scsumrate}
\end{equation}

Let $z=e^{-\frac{x}{a^{\tsc}_K}}$ and $\xi=\exp\left(b^{\tsc}_K/a^{\tsc}_K\right)$. We have $x=-a^{\tsc}_K\ln z$ and $\,dx=-\frac{a^{\tsc}_K}{z}\,dz$. Thus, (\ref{eq:scsumrate}) can be rewritten as
\begin{eqnarray}
C^{\sc}&=&\frac{4}{\ln2}\int_{0}^{1}\frac{1-e^{-z\cdot \xi}}{1-a^{\tsc}_K\ln z}\cdot\frac{a^{\tsc}_K}{z}\,dz,\\
&=&\frac{4}{\ln2}\left[\int_{0}^{\frac{4}{\xi}}\frac{1-e^{-z\cdot \xi}}{\left(1-a^{\tsc}_K\ln z\right)}\frac{a^{\tsc}_K}{z}\,dz+\right.\nonumber\\
&&\left.\int_{\frac{4}{\xi}}^{1}\frac{a^{\tsc}_K\,dz}{\left(1-a^{\tsc}_K\ln z\right)z}\right].\label{eq:slsc}
\end{eqnarray}

The limit of the first term on the right-hand-side (R.H.S) of (\ref{eq:slsc}) becomes negligibly small as $\displaystyle\lim_{K\rightarrow\infty}\frac{4}{\xi}=0$ while the limit of the second term can be computed by exploiting the approximation of $a^{\tsc}_K\approx \rho$ as follows.
\begin{equation}
\lim_{K\rightarrow\infty}\frac{4}{\ln2}\int_{\frac{4}{\xi}}^{1}\frac{a^{\tsc}_K\,dz}{\left(1-a^{\tsc}_K\ln z\right)z}=\lim_{K\rightarrow\infty}4\log \left(b^{\tsc}_K\right).
\end{equation}

Thus, the corresponding scaling law is given by
\begin{equation}
\lim_{K\rightarrow\infty}\frac{C^{\sc}}{4\log\left(b^{\tsc}_K\right)}=1.\label{eq:slsc1}
\end{equation}
In particular, for $\rho=1$, we can approximate $b^{\tsc}_K$ and $a^{\tsc}_K$ as
\begin{eqnarray}
b^{\tsc}_K&\approx&\ln 2K-2\ln\left(1+\ln 2K\right),\label{eq:escbk}\\
a^{\tsc}_K&\approx&1,\label{eq:escak}
\end{eqnarray}
respectively.

Subsequently, the scaling law can be written as follows.
\begin{equation}
\lim_{K\rightarrow\infty}\frac{C^{\sc}_{\rho=1}}{4\log\left(\ln 2K-2\ln\left(1+\ln 2K\right)\right)}=1,\label{eq:slscrho1}
\end{equation}
which stands for a typical scaling law in the $\log\log K$ form.

\subsection{Maximum Ratio Combining (MRC)}
The effective SINR obtained with MRC can be expressed as a ratio of two random variables given by $x=\frac{z}{1/\rho+y}$, where $z$ and $y$ are $\chi^2$ distributed random variables with $2N$ and $2M-2$ degrees of freedom corresponding to the instantaneous signal power of the desired signal and the interfering signal, respectively. In particular, for $M=4$ and $N=2$, we have \cite{Shah00,Rao01}
\begin{equation}\label{eq:mrcpdf}
f^{\mrc}_X(x)=\frac{xe^{-x/\rho}}{\rho^2(1+x)^3}+\frac{6xe^{-x/\rho}}{\rho(1+x)^4}+\frac{12xe^{-x/\rho}}{(1+x)^5}
\end{equation}
and the corresponding CDF can be expressed as
\begin{equation}
F^{\mrc}_X(x)=1-\frac{e^{-x/\rho}}{\left(1+x\right)^3}-\frac{xe^{-x/\rho}}{\rho\left(1+x\right)^3}
-\frac{3xe^{-x/\rho}}{\left(1+x\right)^4}.\label{eq:mrccdf}
\end{equation}

It can be shown that
\begin{equation}
\lim_{x\rightarrow\infty}\frac{1-F_X^{\mrc}(x)}{f_X^{\mrc}(x)}=\rho>0.
\end{equation}
Therefore, the limiting distribution of $\left[F^{\mrc}_X(x)\right]^K$ is also of the Gumbel type.

Following similar steps as in the previous section, we have
\begin{equation}
C^{\mrc}=\frac{4}{\ln2}\int_{0}^{\infty}\frac{1-e^{-e^{-\frac{x}{a^{\tmrc}_K}+
\frac{b^{\tmrc}_K}{a^{\tmrc}_K}}}}{1+x}\,dx.\label{eq:mrcsumrate}
\end{equation}
and
\begin{equation}
\lim_{K\rightarrow\infty}\frac{C^{\mrc}}{4\log\left(b^{\tmrc}_K\right)}=1,\label{eq:slmrc1}
\end{equation}
where $a^{\tmrc}_K$ and $b^{\tmrc}_K$ are the corresponding normalizing factors. In particular for $\rho=1$, we can show that
\begin{eqnarray}
b^{\tmrc}_K&\approx& \ln 3K-2\ln\left(1+\ln K\right),\label{eq:emrcbk}\\
a^{\tmrc}_K&\approx& 1
\end{eqnarray}
and the scaling law has the following $\log\log K$ form.
\begin{equation}
\lim_{K\rightarrow\infty}\frac{C^{\mrc}_{\rho=1}}{4\log\left(\ln 3K-2\ln\left(1+\ln K\right)\right)}=1.\label{eq:slmrcrho1}
\end{equation}

\subsection{Optimal Combining (OC)}
The CDF of the effective SINR obtained OC using $N$ receive antennas in the presence of $M-1$ interfering sources has been derived in \cite{Gao98}. For $M=4$ and $N=2$, the corresponding CDF takes the following form.
\begin{equation}
F_X^{\oc}(x)=1-\frac{e^{-x/\rho}}{\left(1+x\right)^3}-\frac{3xe^{-x/\rho}}{\left(1+x\right)^3}
-\frac{xe^{-x/\rho}}{\rho\left(1+x\right)^3},\label{eq:occdf}
\end{equation}
and the corresponding PDF is
\begin{equation}
f_X^{\oc}(x)=\frac{xe^{-x/\rho}}{\rho^2\left(1+x\right)^4}\left[\left(3\rho+1\right)x
+\left(6\rho^2+6\rho+1\right)\right].
\end{equation}

Since $\displaystyle\lim_{x\rightarrow\infty}\frac{1-F_X^{\oc}(x)}{f_X^{\oc}(x)}=\rho>0$, the limiting distribution of $\left[F^{\oc}_X(x)\right]^K$ is also of the Gumbel type. Similar to the cases of SC and MRC, we can show that
\begin{equation}
C^{\oc}=\frac{4}{\ln2}\int_{0}^{\infty}\frac{1-e^{-e^{-\frac{x}{a^{\toc}_K}+
\frac{b^{\toc}_K}{a^{\toc}_K}}}}{1+x}\,dx.\label{eq:ocsumrate}
\end{equation}
and
\begin{equation}
\lim_{K\rightarrow\infty}\frac{C^{\oc}}{4\log\left(b^{\toc}_K\right)}=1,\label{eq:sloc1}
\end{equation}
where $b^{\toc}_K$ and $b^{\toc}_K$ are the corresponding normalizing factors. In particular for $\rho=1$, we can show that
\begin{eqnarray}
b^{\toc}_K&\approx& \ln 4K-2\ln\ln K,\label{eq:eocbk}\\
a^{\toc}_K&\approx& 1
\end{eqnarray}
and the sum-rate scales like the following $\log\log K$ form.
\begin{equation}
\lim_{K\rightarrow\infty}\frac{C^{\oc}_{\rho=1}}{4\log\left(\ln 4K-2\ln\ln K\right)}=1.\label{eq:slocrho1}
\end{equation}

\section{Simulation Results}

In this section, simulation is performed to confirm our SINR analysis derived in Sec. \ref{sec:analysis}. Unless otherwise specified, we set $M=4$ and $N=2$.

\begin{figure}[h]
\begin{center}
\includegraphics[scale=0.391]{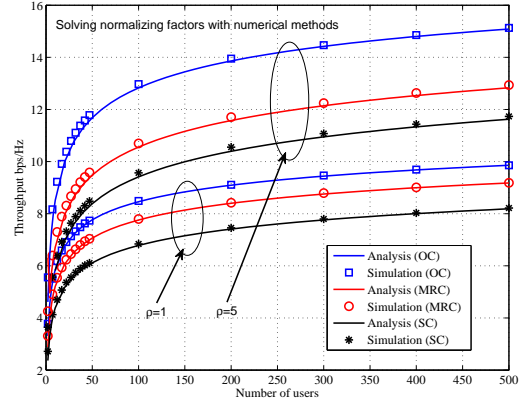}
\caption{Simulation versus analytical results with numerical normalizing factors for $\rho=1,5$.}\label{fig:exact}
\end{center}
\end{figure}

We first compare the asymptotic throughput shown in (\ref{eq:scsumrate}), (\ref{eq:mrcsumrate}) and (\ref{eq:ocsumrate}) against their corresponding simulation results. Figure~\ref{fig:exact} shows the asymptotic throughput curves using the {\em numerical} normalizing factors obtained by numerical methods for $\rho=1$ and $5$. Inspection of Fig. \ref{fig:exact} reveals that the analytical results shown in (\ref{eq:scsumrate}), (\ref{eq:mrcsumrate}) and (\ref{eq:ocsumrate}) are in accord with the simulation results. Despite that the asymptotic analysis is achieved by assuming a large $K$, Fig.~\ref{fig:exact} indicates that the asymptotic analysis is also very accurate for smaller $K$ values. Furthermore, Fig. \ref{fig:exact} confirms that the scheduling scheme with OC can substantially outperform those with MRC and SC whereas the improvement provided by MRC is more apparent in the presence of stronger noise. This is because the scheme with OC is designed to maximize SINR whereas MRC intends to maximize SNR.

\begin{figure}[h]
\begin{center}
\includegraphics[scale=0.391]{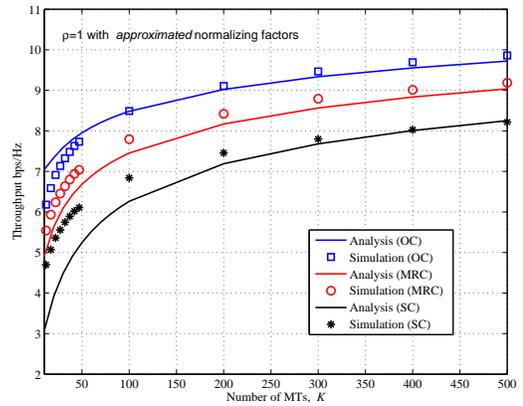}
\caption{Simulation versus analytical results with approximated normalizing factors for $\rho=1$.}\label{fig:approxrho1}
\end{center}
\end{figure}

Next, rather than the numerical solutions, Fig. \ref{fig:approxrho1} depicts the average sum-rates using the {\em approximated} normalizing factors computed in (\ref{eq:escbk}), (\ref{eq:emrcbk}) and (\ref{eq:eocbk}) together with $a_{K}\approx 1$ for $\rho=1$. Since the approximation expressions have been derived by assuming a large $K$, the analytical curves shown in Fig. \ref{fig:approxrho1} approach the simulated curves only when $K$ becomes large.

\begin{figure}[h]
\begin{center}
\includegraphics[scale=0.391]{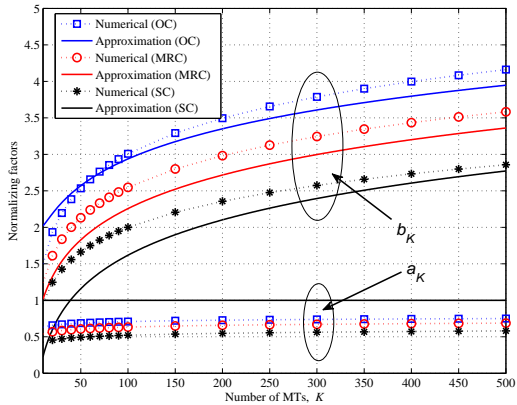}
\caption{Comparison of the numerical and approximated solutions of the normalizing factors for $\rho=1$.}\label{fig:bcomp}
\end{center}
\end{figure}

Finally, to inspect the approximation accuracy of (\ref{eq:escbk}), (\ref{eq:emrcbk}) and (\ref{eq:eocbk}), Fig. \ref{fig:bcomp} shows the numerical and approximated normalizing factors as a function of the number of MTs, $K$. Since solving the exact solutions to the normalizing factors involves the linear-exponential functions, it is in general non-trivial to obtain accurate closed-form expressions for the normalizing factors, which compromises the accuracy of the subsequently derived scaling laws.

\section{Comparison between SIR and SINR analysis}
It is interesting to compare the SINR analysis derived in this work with our previous SIR analysis reported in \cite{Pun07}.

\noindent 1.) On the one hand, it is easy to verify that the CDFs of the effective SINR in (\ref{eq:sccdf}), (\ref{eq:mrccdf}) and (\ref{eq:occdf}) converge the corresponding CDFs of the effective SIR reported in \cite{Pun07} as $\rho$ tends to infinity, respectively. On the other hand, our SIR and SINR-based analysis suggests that
the limiting distributions of the effective SIR and SINR do not belong to the same domain of attraction. Instead, they are of the Frechet-type and Gumbel-type, respectively. It is natural to conjecture that the limiting distribution function of SINR might also converge to the Frechet-type if the noise power becomes zero. However, our results reveal that this intuition is not true. This is because that the limit operator is {\em not} commutative in general.

\noindent 2.) It is generally more difficult to obtain the normalizing factors in the SINR analysis than the SIR analysis since the SINR-based analysis involves exponential-type CDFs and requires solving exponential-linear equations such as (\ref{eq:scbk}). Therefore, it is more computationally advantageous to derive the scaling laws in the SIR-based analysis compared to the SINR-based analysis in the presence of strong interference.

\noindent 3.) When computing the normalizing factors in the SINR-based analysis, we have to carefully take into account the high-order terms in $F_X(x)$. For instance, if the high-order terms in $F_X(x)$ in (\ref{eq:sccdf}), (\ref{eq:mrccdf}) and (\ref{eq:occdf}) are ignored, the resulting simplified CDFs for different schemes will all lead to the same set of normalizing factors, i.e $\frac{e^{-b_K/\rho}}{\left(1+b_K\right)^3}=\frac{1}{K}$. Thus, the performance of OS schemes with different combining techniques cannot be distinguished based on their scaling laws. Since it is generally much easier to compute the normalizing factors with high accuracy in the SIR-based analysis\cite{Pun07}, we argue that the SIR-based scaling laws can better characterize the actual performance of different OS schemes by focusing on the interference-limited scenarios.

\section{Conclusion}
In this paper, we have developed a systematic approach to derive the SINR-based asymptotic throughput and scaling laws for OS schemes by utilizing extreme value theory. In particular, we have investigated the asymptotic throughput and scaling laws of the OS schemes proposed for MIMO-SDMA systems with different linear combining techniques. Our analytical results have shown that the limiting distribution of the effective SINR is of the Gumbel type and the scaling laws follow the $\log\log K$ form. Simulation results have confirmed the effectiveness in improving system throughput by incorporating low-complexity linear combining techniques in OS schemes. Finally, based on the comparison of SIR-based and SINR-based analysis, we have argued that the SIR-based analysis is more advantageous in providing insights into the scheduling performance for SDMA-based systems.

\bibliographystyle{IEEEtranS}
\bibliography{Bib}
\end{document}